\documentclass[11pt]{llncs}
\usepackage{graphicx}
\usepackage{overcite}
\usepackage{times}
\usepackage{fullpage}
\usepackage{footmisc}
\newenvironment{summary}{
\begin{quote} \bf}
{\end{quote}}
\newcounter{lastnote}

\title{Topological network alignment uncovers biological function and phylogeny}
\author{Oleksii Kuchaiev\inst{1}$^{,\ddagger}$ \and Tijana Milenkovi\'{c}\inst{1}$^{,\ddagger}$ \and Vesna Memi\v{s}evi\'{c}\inst{1} \and Wayne Hayes\inst{1,3} \and Nata\v{s}a Pr\v{z}ulj \inst{1,2,}\thanks{To whom correspondence should be addressed; E-mail: natasha@doc.ic.ac.uk .}}
\institute{Department of Computer Science, University of California, Irvine, CA 92697-3435, USA
\and Department of Computing, Imperial College London SW7 2AZ, UK
\and Department of Mathematics, Imperial College London SW7 2AZ, UK
\\ $^\ddagger$These authors contributed equally to this work}
\date{}

\begin{document}

\maketitle

\begin{summary}

Sequence comparison and alignment has had an enormous impact on our
understanding of evolution, biology, and disease.  Comparison and
alignment of biological networks will likely have a similar impact.
Existing network alignments use information external to the networks,
such as sequence, because no good algorithm for purely topological
alignment has yet been devised.  In this paper, we present a novel
algorithm based solely on network topology, that can be used to align
any two networks.  We apply it to biological networks to produce by
far the most complete topological alignments of biological networks to
date.  We demonstrate that both species phylogeny and detailed
biological function of individual proteins can be extracted from our
alignments.  Topology-based alignments have the potential to provide a
completely new, independent source of phylogenetic information.  Our
alignment of the protein-protein interaction networks of two very
different species---yeast and human---indicate that even distant
species share a surprising amount of network topology with each other,
suggesting broad similarities in internal cellular wiring across all
life on Earth.

\end{summary}

\section{Introduction} 

\subsection{Background} 

A network (or graph) is a collection of nodes (or vertices), and
connections between them called edges.  Graphs are used to describe,
model, and analyze an enormous array of
phenomena\cite{Colizza2006,Guimera2007}, including physical systems
like electrical power grids and communication networks, social systems
like networks of friendships or corporate and political hierarchies,
physical relationships such as residue interactions in a folded
protein or software systems such as call graphs or expression and
syntax trees.

A graph $G(V,E)$, or $G$ for brevity, has node set $V$ and edge set
$E$.  The sheer number and diversity of possible graphs
(about $2^{(n^2)}$ of them exist given $n$ nodes) makes graph classification and comparison problems difficult.
One particular comparison problem is called {\it subgraph isomorphism},
which asks if one graph $G$ exists as an exact subgraph of another graph $H(U,F)$.
This problem is {\it NP-complete}, which means that no efficient algorithm is
known for solving it \cite{Cook1971}.
Network alignment\cite{Sharan2006} is the more general problem of
finding the best way to ``fit'' $G$ into $H$ even if $G$ does not
exist as an exact subgraph of $H$.  Some networks, such as the
biological ones that we consider below, may contain noise,
i.e. missing edges, false edges, or both \cite{Venkatesan2009}.
In these cases, and also due to biological variation, it is not even
obvious how to measure the ``goodness'' of an inexact fit.  One
measure could be to assess the number of aligned edges---that is, the
percentage of edges in $E$ that are aligned to edges in $F$.  We call
this the ``edge correctness'' (EC).  However, it is possible for two
alignments to have similar ECs, one of which exposes large, dense,
contiguous, and topologically complex regions that are similar in $G$
and $H$, while the other fails to expose such regions of similarity.
Additionally, although EC can easily be used to measure the quality
of an alignment after the fact, it is not clear how to use it to {\em
direct} an alignment algorithm; in fact, maximizing EC is an NP-hard
problem since it implies solving the subgraph isomorphism problem.
Thus, other strategies must be sought to guide the alignment process.

In the biological context, comparing networks of different organisms
in a meaningful manner is arguably one of the most important problems
in evolutionary and systems biology \cite{Sharan2006}.
Analogous to sequence alignments between genomes, alignments of
biological networks can be useful because we may know a lot about some
of the nodes in one network and almost nothing about topologically
similar nodes in the other network; then, specialized knowledge about
one may tell us something new about the other.
Network alignments can also be used to measure the global similarity
between complete networks of different species.  Given a group of such
biological networks, the matrix of pairwise global network
similarities can be used to infer phylogenetic relationships.

In this paper, we introduce a novel method for the alignment of two
networks that is based {\em solely} on the mutual similarity of their
network topology.  As such, this algorithm could be applied to {\em any}
two networks, not just biological ones.  For example, our algorithm
can be applied to road maps or social networks, which obviously have
no genetic or protein sequence associated with them.  We apply our
method to the alignment of two protein-protein interaction (PPI)
networks and demonstrate that our alignment exposes far more
topologically complex regions of similarity than existing methods
can find.  We also use our method to compute the pairwise all-to-all
network similarity matrix between a group of species, and then
build a phylogenetic tree that bears a striking resemblance to
the one based on sequence comparison.  The significance of these
results are that they extract statistically significant meaning
from a new source of information---pure network topology---that
is independent of sequence or any other commonly used biological
information.  We believe that the results in this paper
just barely scratch the surface of the information that can be
extracted from network topology.

\subsection{Our Approach}

Analogous to sequence alignments, there exist
\emph{local} and \emph{global} network alignments. Thus far, the majority of methods
used for alignment of biological networks have focused on local
alignments \cite{PathBlast, Berg04, Flannick2006,
Liang2006a,Berg2006}.  With local alignments, mappings are chosen
independently for each local region of similarity.  However, local
alignments can be ambiguous, with one node having different pairings
in different local alignments.  In contrast, a global network
alignment provides a unique alignment from every node in the smaller
network to exactly one node in the larger network, even though this
may lead to inoptimal matchings in some local regions.  Local network
alignments are generally not able to identify large subgraphs that
have been conserved during evolution \cite{Berg04,PathBlast}.  Global
network alignment has been studied previously in the context of
biological networks \cite{Singh2007,Flannick2008,GraphM}, but most
existing methods incorporate some {\it a priori} information about
nodes such as sequence similarities of proteins in PPI
networks\cite{Berg2006,Singh2007}, or they use some form of learning
on a set of ``true'' alignments\cite{Flannick2008}.  
In contrast, our alignments are based \emph{solely} on topological
information and do not require learning.  This makes our method
applicable to {\em any} type of network, not just biological ones.

We focus on topology instead of protein sequence because we aim to
discover biological knowledge that is encoded in the PPI network
topology. Since proteins aggregate to perform a function instead of
acting in isolation, analyzing complex wirings around a protein in a
PPI network could give deeper insights into inner working of cells
than analyzing sequences of individual genes. Furthermore, network
topology and protein sequences might give insights into different
slices of biological information and thus, one could loose much
information by focusing on sequence alone.  Although protein sequence similarity
correlates with functional similarity, there exist proteins with
100\% sequence identity that have different functional roles
\cite{Komili2007,Watson2005,Whisstock2003}.  Thus, restricting analysis
to sequences might give incorrect functional assignments.  Similarly,
although high protein sequence similarity correlates with
similarity in 3-dimensional structure, sequence-similar
proteins can have structures that differ significantly from one
another \cite{Kosloff2008}. Thus, sequence-based homology analyses may
mask important structural and functional information. On the other
hand, since the structure of a protein is expected to define the
number and type of its potential interacting parters in the PPI
network, sequence-similar but structurally-dissimilar proteins are
expected to have different PPI network topological
characteristics. Moreover, entirely different sequences can produce
identical structures \cite{Laurents1994,Whisstock2003}. In cases where
such proteins are expected to share a common function, sequence-based
function prediction would fail, where network topology-based one
would not. Finally, we show that both sequence and topology have
similar predictive power with respect to Gene Ontology (GO) terms\cite{Go00}
(Supplementary Figure 1), demonstrating that network topology can
provide as much functional information as protein sequences.
Since our goal is to uncover biological knowledge encoded in the
topology of PPI networks, our alignments do not use protein sequence
information. Thus, our method can align {\em any} type of network, not
just biological ones.  Note, however, that inclusion of sequence
component into the cost function of our method is trivial (see
Section \ref{sec:meth}), but this is out of the scope of the manuscript.

Obviously, if one is to build meaningful alignments based solely upon
network topology, one must first have a highly constraining {\em
measure} of topological similarity.  The simplest (and weakest)
description of the topology of a node is its {\it degree}, which is
the number of edges that touch it. Our much more highly constraining
measure is a generalization of the degree of a node.  We define a {\it
graphlet} as a small, connected, \emph{induced} subgraph of a larger
network \cite{Przulj04,Przulj06,Przulj2006}. An \emph{induced}
subgraph on a node set $X \subseteq V$ of $G$ is obtained by taking
$X$ and \emph{all} edges of $G$ having both end-nodes in $X$.  Figure
\ref{fig:graphlets} shows all the graphlets on 2, 3, 4, and 5 nodes.
For a particular node $v$ in a large network, we define a vector of
``graphlet degrees''\cite{Milenkovic2008} that counts the number of
each kind of graphlet that touch $v$ (Figure
\ref{fig:sig_il}).  This vector, or {\it signature}, of $v$ describes
the topology of its neighborhood and captures its interconnectivities
out to a distance of 4 (see Section \ref{sec:meth1} and Figure
\ref{fig:sig_il})\cite{Milenkovic2008}.  This measure is superior to
all previous measures, since it is based on all up to 5-node
graphlets, which is practically enough due to the small-world nature
of many real-world networks \cite{Watts-Strogatz98}.

For our purposes, an alignment of two networks $G$ and $H$ consists of
a set of ordered pairs $(x,y)$, where $x$ is a node in $G$ and $y$ is
a node in $H$.  Our algorithm, called GRAAL (GRAph ALigner),
incorporates facets of both local and global alignment.  We match
pairs of nodes originating in different networks based on their {\it
signature similarity}\cite{Milenkovic2008} (see Section \ref{sec:meth1}),
where a higher signature similarity between two nodes corresponds
to a higher topological similarity between their extended neighborhoods (out
to distance 4). 
The cost of aligning two nodes is modified to align the densest parts
of the networks first; the cost is reduced as the degrees of both
nodes increase, since higher degree nodes with similar signatures
provide a tighter constraint than correspondingly similar low degree
nodes (see Section \ref{sec:meth2} and the Supplementary Information); $\alpha$ is a
parameter in [0,1] that controls the contribution of the node
signature similarity to the cost function, the other contribution
being simply the degree of the node (see Section \ref{sec:meth2}).
In the case of two node
alignments comparing equally, the tie is broken randomly.  Thus,
different runs of the alignment algorithm can produce different
results, although we generally find that a deterministic ``core''
alignment remains across all runs.

We align each node in the smaller network to exactly one node in the
larger network.  The matching proceeds using a technique analogous to
the ``seed and extend'' approach of the popular BLAST\cite{Altschul90}
algorithm for sequence alignment: we first choose a single ``seed''
pair of nodes (one node from each network) with high signature
similarity.  We then expand the alignment radially outward around the
seed as far as practical using a greedy algorithm (see Section \ref{sec:meth2}).
Although local in nature, our algorithm produces large and dense
global alignments.  By ``dense'' we mean that the aligned subgraphs
share many edges, which would not be the case in a low-quality or
random alignment.  We believe that the high quality of our alignments
is based less on the details of the extension algorithm and more on
having a good measure of pair-wise topological similarity between
nodes\cite{Milenkovic2008}.

\section{Results and Discussion} 

\subsection{Pairwise Alignment of Yeast and Human PPI Networks}\label{sect:pairwise} 

Using GRAAL, we align the human PPI network of Radivojac et
al. \cite{Radivojac2008} to the Collins et al. yeast PPI network
\cite{Krogan2007}, which we call ``human1'' and ``yeast2,''
respectively.  We chose yeast as our second species because currently
it has a high quality PPI network, with 16,127 interactions (edges)
among 2,390 proteins (nodes).
The ``best'' alignment (defined below) found by GRAAL aligns 1,890 of
the edges in yeast2 to edges in human1.  Thus, the edge correctness
(EC) of our alignment is 11.72\%.  There are 970 nodes involved in
these ``correct'' edge alignments, representing 40\% of all yeast2
nodes.  We obtained similar EC
for aligning other yeast \cite{Krogan2007,BIOGRID} and human
\cite{BIOGRID,HPRD,Rual05} networks (Supplementary Figure 2).
The best alignment is defined as follows. Due to the existence of the
$\alpha$ parameter in the cost function (as explained above) and some
randomness in the GRAAL algorithm (see Section \ref{sec:meth2} and the Supplementary
Information for details), the actual alignments and ECs vary across
different values of $\alpha$, and across different runs of the algorithm for
the same $\alpha$. With this in mind, the best alignment is the
alignment with the highest EC over all values of $\alpha$, and over all runs
for the given $\alpha$. The highest EC is obtained for $\alpha$ of
0.8; the minimum EC over all runs for this $\alpha$ is higher than the
maximum EC over all runs for any other $\alpha$. Thus, we focus on
alignments produced for $\alpha$ of 0.8. Variation of EC over
different runs for this $\alpha$ is small, with minimum and maximum EC
of 11.5\% and 11.72\%, respectively. Moreover, intersection of
alignments from
up to 40 different runs at $\alpha$ of 0.8 contains 1,433 pairs,
i.e., about $60\%$ of the entire alignment. We call this
intersection the {\it core} alignment.

In addition to counting aligned edges, it is important that the
aligned edges cluster together to form large and dense connected
subgraphs, in order to uncover such regions of similar topology.  We
define a {\it common connected subgraph} (CCS) as a connected subgraph
(not necessarily induced) that appears in both networks.  
The largest CCS in our best alignment 
(Figure \ref{fig:11}A) has 900 interactions amongst 267 proteins,
which comprises 11.2\% of the proteins in the yeast2 network.  Our
second largest CCS has 286 interactions amongst 52 nodes, depicted in
Figure \ref{fig:11}B.  The entire common subgraph is presented in
Supplementary Figure 3.

\subsection{Comparison with Other Methods}

GRAAL uncovers CCSs that are substantially larger and denser than
those produced by currently published algorithms.  The best currently
published global alignment of similar networks is the alignment of
yeast and fly by IsoRank \cite{Singh2007}, which uses sequence
information in addition to topological information.  It aligns 1,420
edges, but its largest CCS contains just 35 nodes and 35 edges.  Our
largest CCS aligns 25.7 times as many edges and 7.6 times as many
nodes in human-yeast than IsoRank does in fly-yeast.  Our {\em second}
largest CCS has a similar number of nodes to IsoRank's {\em largest},
but is 8.2 times denser in terms of edges.  Furthermore, we applied
IsoRank to our yeast2-human1 data using only topological information.
We found that it aligns 628 interactions (giving an edge correctness
of only 3.89\%), with its largest CCS having just 261 interactions
among 116 proteins.
Recently, IsoRankN, an algorithm for global alignment of
\emph{multiple} networks, has been introduced\cite{IsoRankN}. However, 
a comparison with GRAAL is not feasible, since the alignment output of
the two algorithms is different.  While GRAAL's output is a list of
one-to-one node mappings between the networks being aligned,
IsoRankN's alignment contains sets of network-aligned proteins, where
no two sets overlap, but each set can contain more than one node
(i.e., many-to-many node mapping) from each of the networks being
aligned; thus, IsoRankN's output can not be quantified topologically
with EC.
Another popular global network alignment method is
Graemlin\cite{Flannick2008}. We do not compare our alignment to one
produced by Graemlin because Graemlin requires a variety of other
input information, including phylogenetic relationships between the
species being aligned.  In contrast, GRAAL's {\em output} can be used
to infer phylogenetic relationships.
Finally, other methods potentially better than IsoRank
exist\cite{GraphM}; however, their current implementations failed to
process networks of the size of yeast2 and human1\footnote{Personal
communication with the authors.}.

\subsection{Statistical Significance of GRAAL's Yeast-Human Alignment}

In the following three paragraphs, we look at three distinct ways in
which to judge the statistical significance of our alignment: first,
we judge the quality of our alignment compared to a random alignment
of these two particular networks; second, we comment on the amount of
similarity found between yeast and human in our alignment; and third,
we interpret the biological significance of our alignment.  Section \ref{sec:meth} and Supplementary Information provide more details on 
all of the above.

Given a random alignment of yeast2 to human1, the probability of
obtaining an edge correctness of $11.72\%$ or better ($p$-value) is
less than $7\times 10^{-8}$.  The probability of obtaining a large CCS
would be significantly smaller, so this represents a weak upper bound
on our $p$-value.

Judging the amount of similarity found between the yeast2 and human1
networks in our alignment requires us to state carefully what we are
comparing against.  If we align 
with GRAAL
networks drawn from several different
random graph models\cite{GraphCrunch} that have the same number of
nodes and edges as yeast2 and human1, we find that the edge
correctness between random networks is significantly lower than the
edge correctness of our yeast2-human1 alignment.  For example,
aligning two Erd\"{o}s-R\'{e}nyi random graphs with the same degree
distribution as the data (``ER-DD'') gives an edge correctness of only
about $0.31\pm 0.22\%$.  Similar alignments of Barab\'{a}si-Albert
type scale-free networks (``SF-BA'')
\cite{Barabasi99}, stickiness model networks (``STICKY'') \cite{PrzuljHigham06}, or
$3$-dimensional geometric random graphs (``GEO-3D'') \cite{Przulj04},
give edge correctness scores of only $2.86\pm 0.57\%$, $5.89\pm
0.39\%$ and $8.8\pm 0.39\%$, respectively.  Accepting GEO-3D as the
best available null model (see Section \ref{sec:meth3}), the $p$-value of our
yeast2-human1 alignment is at most $8.4\times 10^{-3}$.  This tells us
that yeast and human, two very different species, enjoy more network
similarity than chance would allow.

We measure the biological significance of our alignment by counting
how many of our aligned pairs share common Gene Ontology (GO)
terms\cite{Go00}.  GO terms succinctly describe the many biological
properties that a given protein may have. For this analysis, we
consider the ``complete'' GO annotation data set, containing all GO
annotations, independent of GO evidence code.  GO annotation data was
downloaded in September 2009. Across our
entire best yeast2-human1 alignment,
%
45.1\%, 15.6\%, 5.1\%, and 2.0\% of aligned protein pairs share at
least one, two, three, and four GO terms, respectively.  Compared to
random alignments, the $p$-values for these percentages are all in the
$10^{-6}$ to $10^{-8}$ range.  Furthermore, the results improve across
GRAAL's {\it core} yeast2-human1 alignment:
50.9\%, 19.3\%, 7.3\%, and 3.0\% of aligned protein pairs share at
least one, two, three, and four GO terms, respectively; the p-values for
these percentages are all in the $10^{-8}$ to $10^{-9}$ range.  Our
results are better then those achieved by IsoRank. In the global
alignment produced by IsoRank 44.2\%, 14.1\%, 4.1\%, and 1.5\% of
aligned protein pairs share at least one, two, three, and four GO terms
in common, respectively.
Similarly, if we restrict our analysis only to the largest CCS, in 
GRAAL's CCS, the percentages are 67.2\%, 22.0\%, and 5.2\% for sharing
at least 1, 2, and 3 common GO terms, respectively, while in IsoRank's
CCS, these percentages are only 60.6\%, 11.9\%, and 0\%, respectively.

\subsection{Application to Protein Function Prediction}

With the above validations in hand, we believe that GRAAL's alignments
can be used to predict biological characteristics (i.e., GO molecular
function (MF), biological process (BP), and cellular component (CC))
of un-annotated proteins based on their alignments with annotated
ones.

Here, we distinguish between two different sets of GO annotation data:
the complete set described above, containing all GO annotations,
independent of GO evidence codes, and ``bio-based'' set, containing GO
annotations obtained by experimental evidence codes only (see
\cite{Go00} for details).  Since in the complete GO annotation data set, many GO
terms were assigned to proteins computationally (e.g., from sequence
alignments), this set is biologically less confident than the
bio-based one. We make predictions with respect to both GO annotation
data sets, as described below.

First, we analyze GRAAL's best yeast2-human1 alignment (i.e., the
alignment with the highest EC over all runs for alpha of 0.8, as
explained in Section \ref{sect:pairwise}) to identify protein pairs
where one of the proteins is annotated with a ``root'' GO term:
GO:0003674 for MF, GO:0008150 for BP, or GO:0005575 for CC, signifying
that a protein is expected to have a MF, BP, or CC, respectively, but
that no information was available as of the date of
annotation\cite{Go00}.  Next, we check if aligned partners of such
proteins are annotated with a known MF, BP, or CC GO term,
correspondingly, with respect to both the complete and bio-based GO
annotation data sets. If so, we assign all known MF, BP, or CC GO
terms to the protein currently annotated by the corresponding ``root''
GO term.

With respect to the complete GO data set, we predict MF for 44 human
and 435 yeast proteins, BP for 53 human and 157 yeast proteins, and CC
for 52 human and 54 yeast proteins. Since GO database offers a list
with an explicit note that a protein is not associated with a given GO
term, we were able to examine directly whether our predictions
contradicted this list. We found no contradictions in GO database for
any of the yeast or human proteins with respect to MF or BP; we found
contradiction only for one of our human predictions with respect to
CC. We also attempted to validate 
all of our predictions using the literature search and text mining
tool CiteXplorer\cite{EBIWebservice}. For 34.1\%, 43.4\%, and 46.2\%
of our MF, BP, and CC human predictions, respectively, this tool found
at least one article mentioning the protein of interest in the context
of at least one of our predictions for that protein.  For yeast, these
percentages are 42.07\%, 3.18\%, and 12.96\%, respectively.
Our human and yeast predictions made with respect to the complete GO
data set are presented in Supplementary Tables 1 and 2, respectively.

With respect to bio-based GO data set, we predict MF for 30 human and
214 yeast proteins, BP for 42 human and 41 yeast proteins, and CC for
45 human and 17 yeast proteins. None of these predictions were
contradicted in the GO database. We validated with CiteXplorer 10\%,
4.76\%, and 20\% of our bio-based MF, BP, and CC human predictions,
respectively. We also validated 48.1\% of our bio-based MF yeast
predictions. 
Our human and yeast predictions made with respect to bio-based GO data
set are presented in Supplementary Tables 3 and 4, respectively.

\subsection{Reconstruction of Phylogenetic Trees by Aligning Metabolic Pathways Across Species}

Finally, we describe a completely different application: how purely
topological alignment of metabolic networks obtained by GRAAL can be
used to recover phylogenetic relationships.

Several studies analyzing metabolic pathways in different species have
aimed to find an evolutionary relationship between those species and
construct their phylogenetic trees
\cite{Forst2001,Heymans2003,Zhang2006,Suthram2005b}.  Different distance metrics have 
been used for constructing phylogenetic trees. For example,
similarities between pathways have been computed from sequence
similarities between corresponding substrates and enzymes from
individual pathways\cite{Forst2001} or as a combination of
similarities of enzymes from individual metabolic networks and
topologies of these networks\cite{Heymans2003,Suthram2005b}. The
similarity of enzymes is based on the similarity of their sequences,
structures, or Enzyme Commission numbers \cite{Enz}. The topological similarity
of two pathways has been based on the similarity between nodes
(corresponding to enzymes) and the similarity of their neighborhoods,
measuring whether a node influences similar nodes and whether it is
influenced by similar nodes itself\cite{Heymans2003}.  In addition,
topological similarity of metabolic pathways combining global network
properties, such as the diameter and clustering coefficient, and
similarities of shared node (i.e., enzyme) neighborhoods has been
used\cite{Zhang2006}. 

Therefore, although related attempts exist\cite{Suthram2005b}, they all still use
some biological or functional information such as sequence
similarities to define node similarities and derive phylogenetic
trees from pathways. 
Since we use only network topology to define protein similarity, our
information source is fundamentally different.  Thus, our algorithm
recovers phylogenetic relationships (but not the evolutionary
timescale of species divergence) in a completely novel and independent
way from all existing methods for phylogenetic recovery.

It has been shown that PPI network structure has subtle effects on the
evolution of proteins and that reasonable phylogenetic inference can
only be done between closely related species\cite{Stumpf05a}.  In the
KEGG pathway database, there are 17 Eucaryotic organisms with fully
sequenced genomes \cite{Kanehisa2000}, of which seven are protists,
six are fungi, two are plants, and two are animals.  Here we focus on
protists (see the Supplementary Information for fungi).  For each
organism, we extract the union of all metabolic pathways from KEGG,
and then find all-to-all pairwise network alignments between species
using GRAAL.  The edge correctness scores between pairs of protist
networks range from 29.6\% to 76.7\%.  We create phylogenetic trees
using the average distance
algorithm\footnote{http://www.mathworks.com/access/helpdesk/help/toolbox/bioinfo/index.html},
with pairwise edge correctness as the distance measure.  We compare
our phylogenetic trees to the published
ones\footnote{http://fungal.genome.duke.edu/} obtained from
genetic or amino acid sequence alignments
\cite{Pennisi2003,Keeling2000}.
Figure \ref{fig:protists_tree} presents our phylogenetic tree for
protists and shows that it is very similar to that found by sequence
comparison\cite{Pennisi2003}.  We can estimate the statistical
significance of our tree by measuring how it compares to trees built
from random networks of the same size as the metabolic networks (see
the Supplementary Information); we find that the $p$-value of our tree
is less than $1.3\times10^{-3}$. Phylogenetic trees based on
alignments made by IsoRank do not differ significantly from random
ones (see the Supplementary Information). We also find that the
topologies of the entire metabolic networks of Cryptosporidium parvum
and Cryptosporidium hominis are very similar, having edge correctness
of 75.72\%. This result is encouraging since these organisms are two
morphologically identical species of Apicomplexan protozoa with 97\%
genetic sequence identity, but with strikingly different hosts
\cite{Tanriverdi2006} that contribute to their
divergence \cite{Xu2004}.

Note that all of the metabolic networks that we align are derived from
a mix of both experimental data and genetic sequence-based data.
Thus, the fact that we recover almost the same tree as sequence-based
methods is a strong validation of our method.  Once the KEGG database
gets updated to have metabolic pathways that are determined solely by
experiment, our phylogenetic trees will provide a new and completely
independent, objective source of phylogenetic information, as well as
a novel, independent verification of the sequence-based phylogeny.
Given that our phylogenetic tree is slightly different from that
produced by sequence, there is no reason to believe that the
sequence-based one should {\it a priori} be considered the correct
one.  Sequence-based phylogenetic trees are built based on multiple
alignment of gene sequences and whole genome alignments.  Multiple
alignments can be misleading due to gene rearrangements, inversions,
transpositions, and translocations that occur at the substring level.
Furthermore, different species might have an unequal number of genes
or genomes of vastly different lengths.  Whole genome phylogenetic
analyses can also be misleading due to non-contiguous copies of a gene
or non-decisive gene order\cite{Otu2003}.  Finally, the trees are
built incrementally from smaller pieces that are ``patched'' together
probabilistically\cite{Pennisi2003}, so probabilistic errors in the
tree are expected.  Our tree suffers from none of the above problems,
although it may suffer from others that are presumably independent of
those above.

\section{Conclusions}

Network alignment has applications across an enormous span of domains,
from social networks to software call graphs.  In the biological
domain, the mass of currently available network data will only
continue to increase and we believe that high-quality topological
alignments can yield new and pivotal insights into function,
evolution, and disease.

\section{Methods}\label{sec:meth}

A graph $G(V,E)$, or $G$ for brevity, has node set $V$ and edge set
$E$.  Given $n=|V|$ nodes, the maximum number of undirected edges is
$M=n(n-1)/2$, and the number of possible undirected graphs on $n$
nodes is thus $2^M$.  The sheer number and diversity of possible
graphs makes graph classification and comparison problems difficult.
One of those problems is called {\it subgraph isomorphism}: given two
arbitrary graphs $G(V,E)$ and $H(U,F)$ such that $|V|\leq|U|$, does
$G$ exist as a subgraph of $H$?  That is, is there a discrete map
$\sigma : V \rightarrow U$ defined $\forall v\in V$ such that
$(x,y)\in E \Rightarrow (\sigma x, \sigma y)\in F$?  This problem is
{\it NP-complete}, which means that no efficient algorithm is known
for finding the mapping $\sigma$---the only known generally applicable
way is to search through all possible mappings from $V$ to
$U$\cite{Cook1971}.  Since the number of such mappings is exponential
in both $|V|$ and $|U|$, this is considered
an intractable problem.

\subsection{Graphlet Degree Signatures and Signature Similarities}\label{sec:meth1}

\begin{figure}[h]
\begin{center}
\includegraphics[scale=0.65]{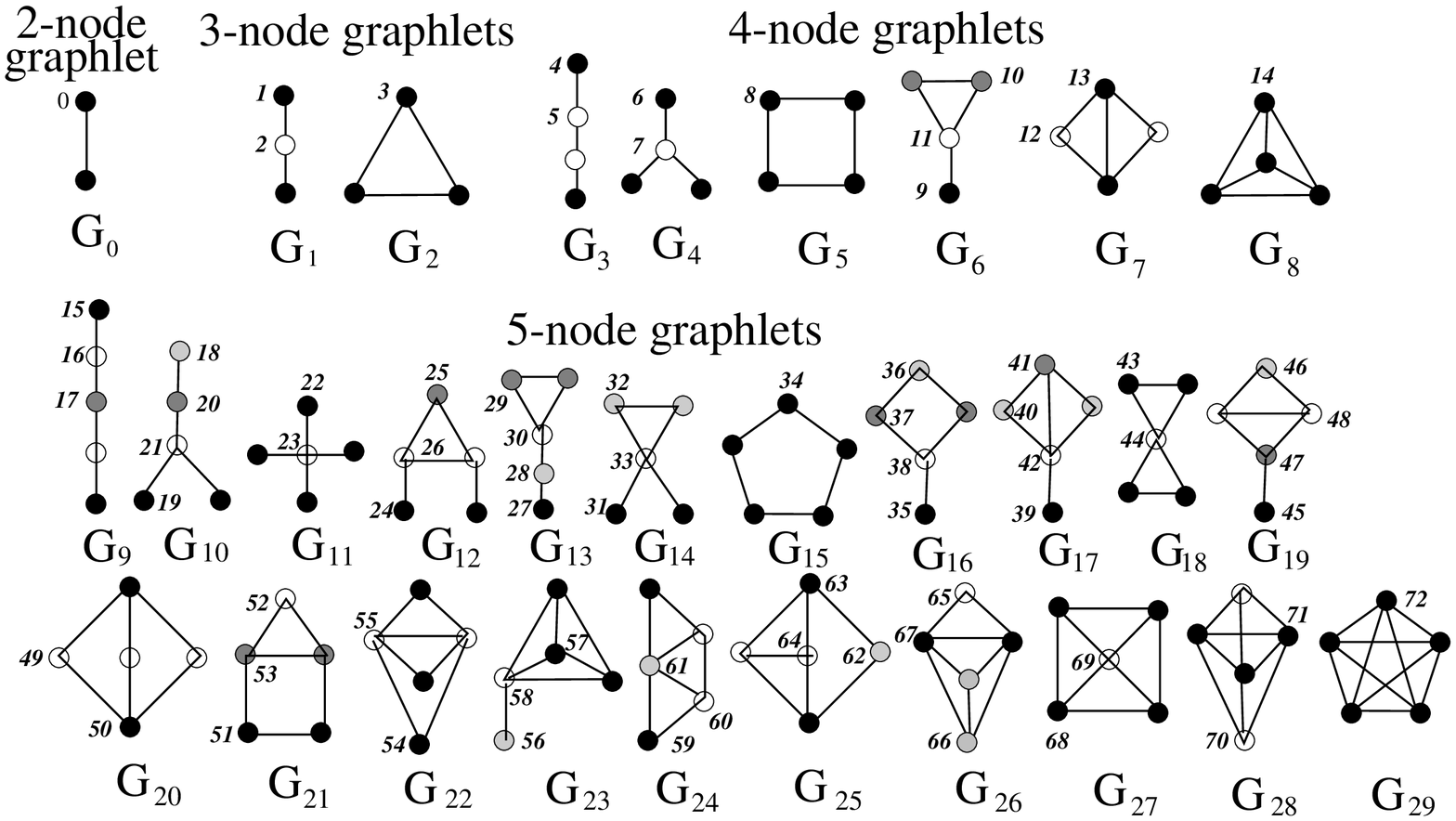}
\caption{All the connected graphs on up to 5 nodes.
When appearing as an induced subgraph of a larger graph, we call them
{\it graphlets}.  They contain 73 topologically unique node types,
called ``automorphism orbits.''  In a particular graphlet, nodes
belonging to the same orbit are of the same shade.  Graphlet $G_0$ is
just an edge, and the degree of a node historically defines how many
edges it touches.  We generalize the degree to a 73-component ``graphlet degree'' vector
that counts how many times a node is touched by each particular
automorphism orbit\cite{Przulj2006}.  }
\label{fig:graphlets}
\end{center}
\end{figure}

\begin{figure}[h]
\begin{center}
\includegraphics[scale=0.8]{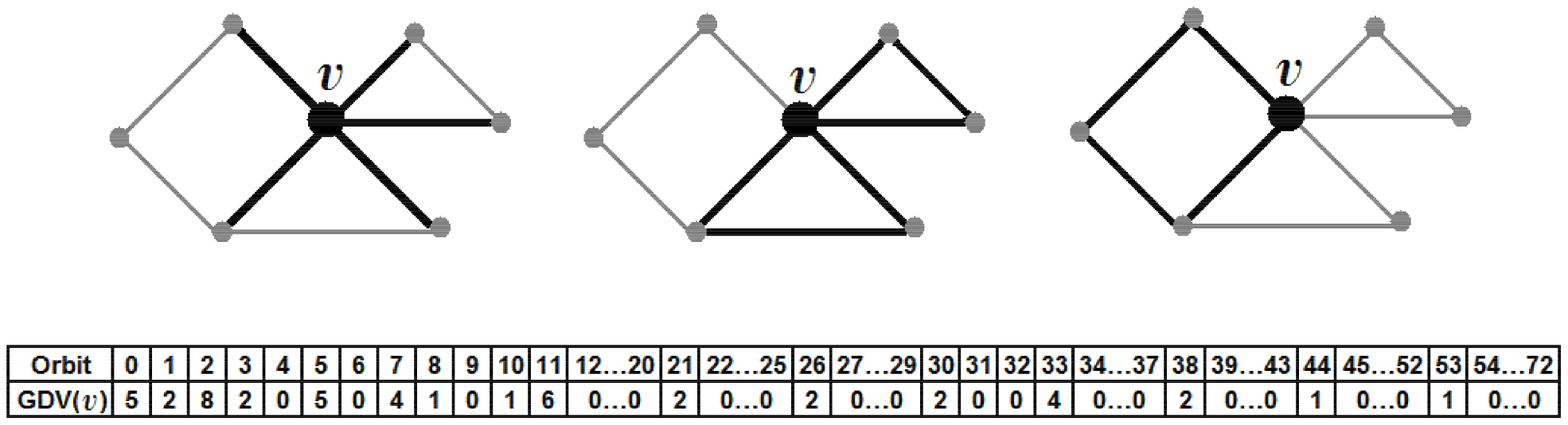}\\
\end{center}
\caption{An illustration of how the degree of  node $v$ in
the leftmost panel is generalized into its ``graphlet degree vector,''
or ``signature,'' that counts the number of different graphlets that
the node touches, such as triangles (middle panel) or squares
(rightmost panel).  Values of the 73 coordinates of the graphlet
degree vector of node $v$, $GDV(v)$, are presented in the table.
}
\label{fig:sig_il}
\end{figure}

\begin{figure}[ht]
\textsf{(A)}
\includegraphics[scale=0.50]{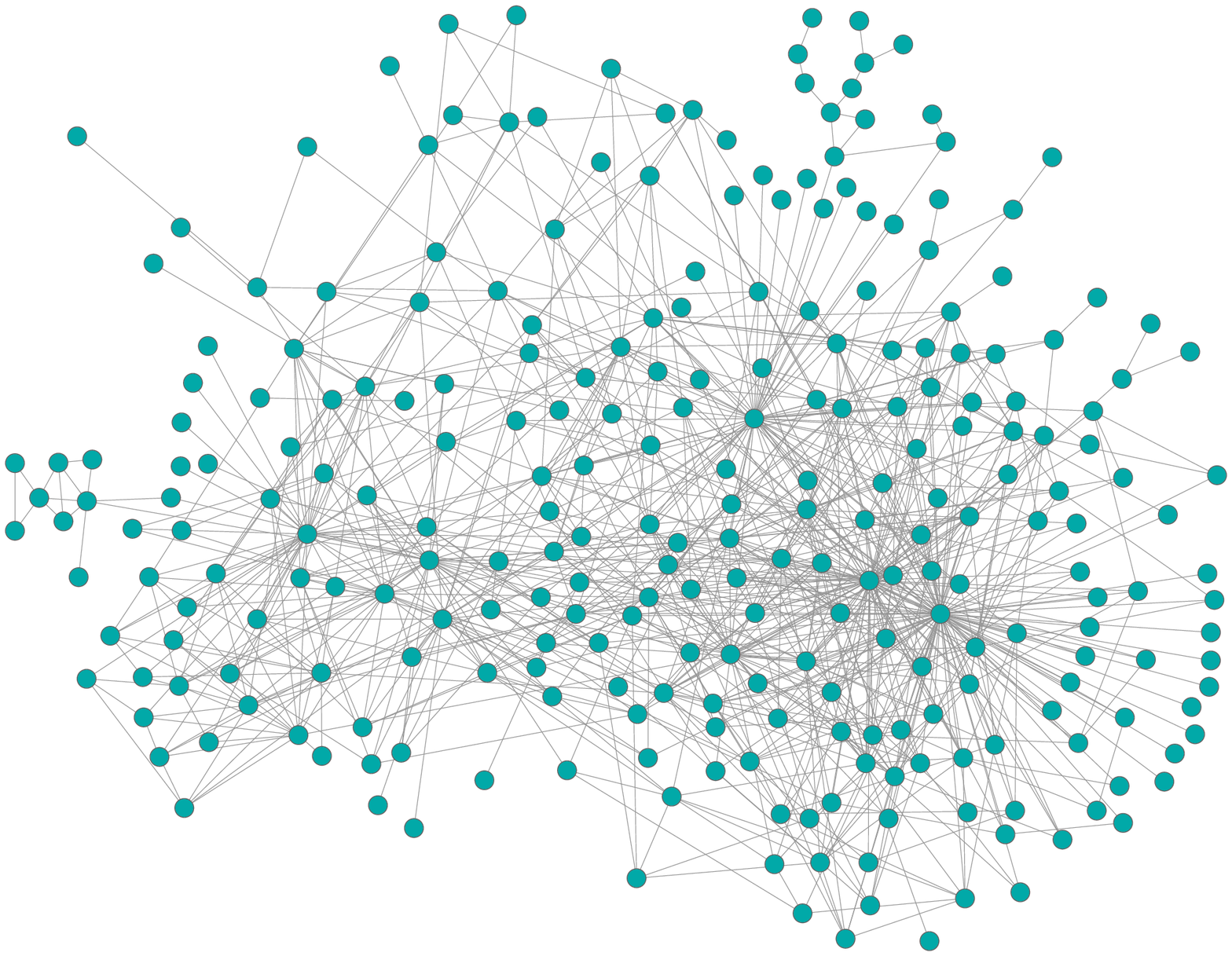}
\\
\textsf{(B)}
\includegraphics[scale=0.5]{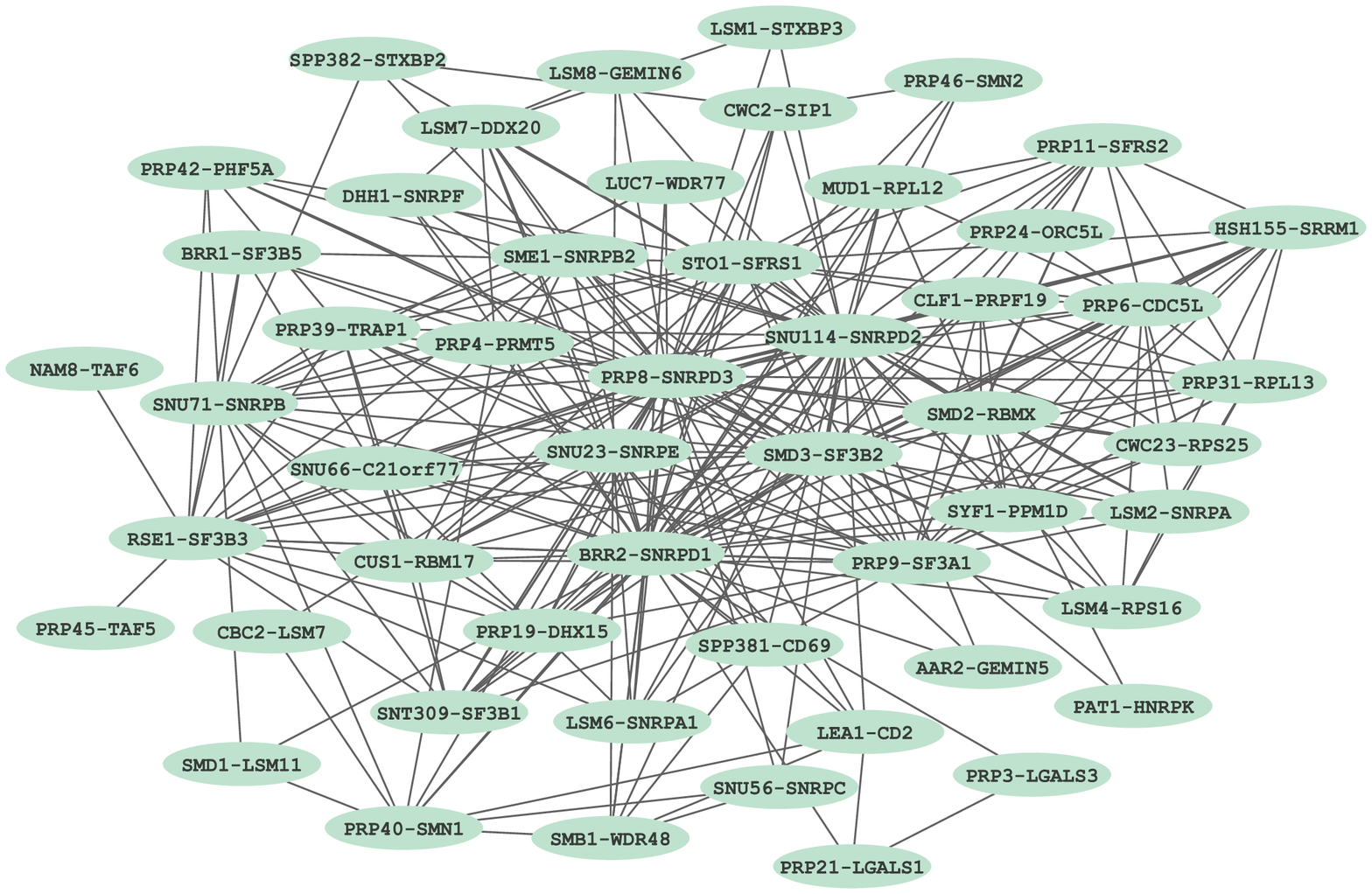}
\caption{The alignment of yeast2 and human1 PPI networks.
An edge between two nodes means that an interaction exists in both species
between the corresponding protein pairs.  Thus, the displayed networks
appear, in their entirety, in the PPI networks of both species.
(A) The largest {\it common connected subgraph} (CCS) consisting of 900 interactions amongst 267 proteins.
(B) The second largest CCS consisting of 286
interactions amongst 52 proteins;
each node contains a label denoting a pair of yeast and human proteins
that are aligned.  }
\label{fig:11}
\end{figure}

\begin{figure}[ht]
\begin{center}
\includegraphics[scale=0.7]{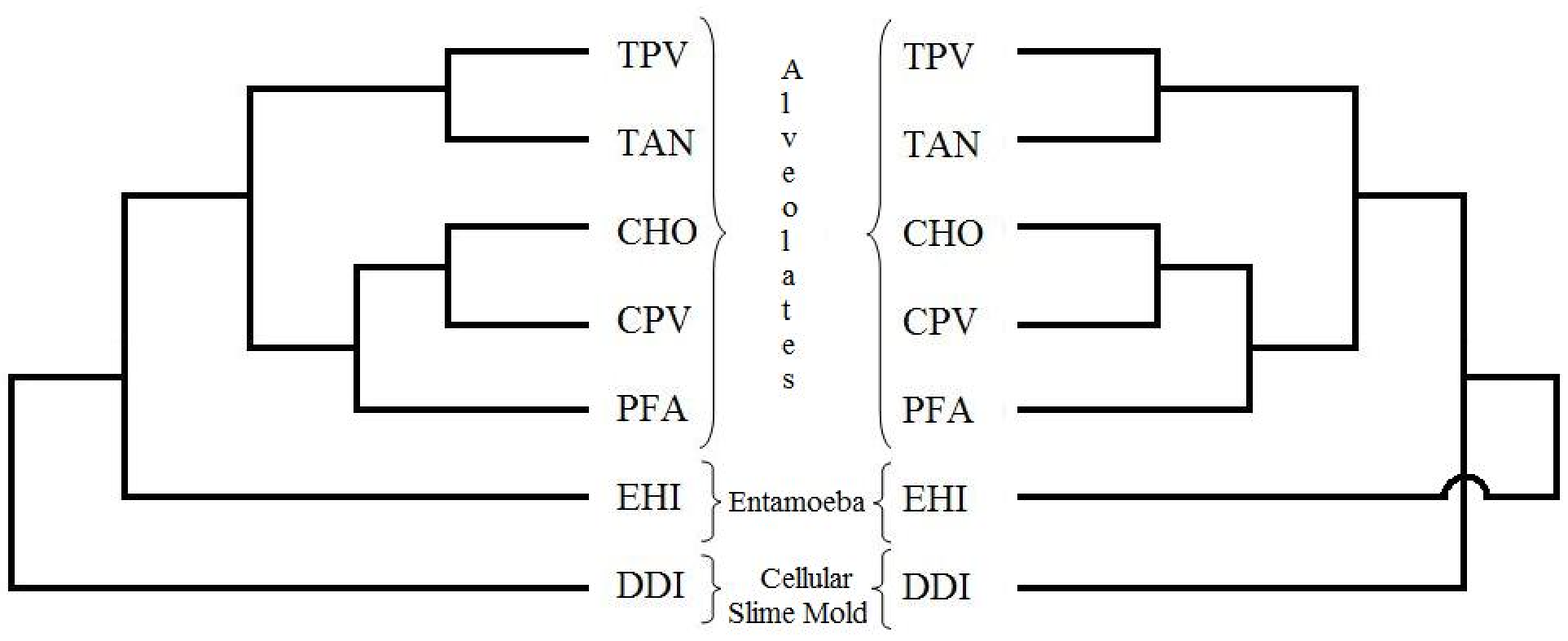}
\end{center}
\caption{
Comparison of the phylogenetic trees for protists obtained by genetic
sequence alignments and by GRAAL's metabolic network alignments. Left:
The tree obtained from genetic sequence
comparison\cite{Pennisi2003}. Right: The tree obtained from GRAAL. The
following abbreviations are used for species: CHO - Cryptosporidium
hominis, DDI - Dictyostelium discoideum, CPV - Cryptosporidium parvum,
PFA - Plasmodium falciparum, EHI - Entamoeba histolytica, TAN -
Theileria annulata, TPV - Theileria parva. The species are grouped
into the following classes: ``Alveolates,'' ``Entamoeba,'' and
``Cellular Slime mold.''}
\label{fig:protists_tree}
\end{figure}

GRAAL aligns a pair of nodes originating in different networks based
on a similarity measure of their local neighborhoods
\cite{Milenkovic2008}.  This measure  generalizes
the degree of a node, which counts the number of edges that the node
touches, into the vector of \emph{graphlet degrees}, counting the
number of graphlets that the node touches, for all 2-5-node graphlets
(see Figure \ref{fig:graphlets}).  Note that the degree of a node is
the first coordinate in this vector, since an edge (graphlet $G_0$ in
Figure \ref{fig:graphlets}) is the only 2-node graphlet.  Since it is
topologically relevant to distinguish between, for example, nodes
touching graphlet $G_1$ at an end or at the middle, the notion of
\emph{automorphism orbits} (or just
\emph{orbits}, for brevity) is used.
By taking into account the ``symmetries'' between nodes of a graphlet,
there are 73 different orbits across all 2- to 5-node
graphlets.  We number the orbits from 0 to 72\cite{Przulj2006}.
The full vector of 73 coordinates is the {\it signature} of a node
(Figure \ref{fig:sig_il}).

The signature of a node provides a novel and highly constraining
measure of local topology in its vicinity and comparing the signatures
of two nodes provides a highly constraining measure of local
topological similarity between them.  The {\it signature
similarity}\cite{Milenkovic2008} is computed as follows.  For a node
$u\in G$, $u_i$ denotes the $i^{th}$ coordinate of its signature
vector, i.e., $u_{i}$ is the number of times node $u$ is touched by an
orbit $i$ in $G$. The distance $D_{i}(u,v)$ between the $i^{th}$
orbits of nodes $u$ and $v$ is defined as $D_i(u,v) = w_i \times
\frac{|log(u_i + 1) - log(v_i + 1)|}{log(max\{u_i, v_i\} + 2)}$, where
$w_{i}$ is a weight of orbit $i$ that accounts for dependencies
between orbits; for example, differences in counts of orbit 3 will
imply differences in counts of all orbits that contain a triangle,
such as orbits 10-14, 25, 26, etc. and thus, a higher weight is
assigned to orbit 3, $w_3$, than to the orbits that contain
it\cite{Milenkovic2008}.
The total distance $D(u,v)$ between nodes $u$ and $v$ is defined as:
$D(u,v) = \frac{\sum_{i=0}^{72}D_i}{\sum_{i=0}^{72}w_i}$.  The
distance $D(u,v)$ is in [0, 1), where distance 0 means that signatures
of nodes $u$ and $v$ are identical.  Finally, the signature
similarity, $S(u,v)$, between nodes $u$ and $v$ is $S(u,v) = 1 -
D(u,v)$.

\subsection{GRAAL (GRAph ALigner) Algorithm}\label{sec:meth2}

When aligning two graphs $G(V,E)$ and $H(U,F)$, GRAAL first computes
costs of aligning each node in $G$ with each node in $H$.  The cost of
aligning two nodes takes into account the signature similarity between
them, modified to reduce the cost as the degrees of both nodes
increase, since higher degree nodes with similar signatures provide a
tighter constraint than correspondingly similar low degree nodes (see
the Supplementary Information).
$\alpha$ is the parameter in [0,1] that controls the contribution of
the signature similarity to the cost function; that is, $1-\alpha$ is
the parameter that controls the contribution of node degrees to the
cost function. In this way, we align the densest parts of the networks
first.  

It is also possible to add protein sequence component to the cost
function, to balance between topological and sequence similarity of
aligned nodes.  This can be done trivially by adding another parameter
$\beta$ to the cost function that would control the contribution of
the current topologically-derived costs, while $1-\beta$ would control
the contribution of node sequence similarities to the total cost
function; similar has been done by other relevant studies
\cite{Singh2007,IsoRankN,GraphM}.  However, as we aim to extract only
biological information encoded in network topology, analyzing how
balancing between the topological and sequence similarity affects the
resulting alignments is out of the scope of our manuscript and is the
subject of future work.

GRAAL chooses as the initial seed a pair of nodes $(v,u)$, $v\in V$
and $u\in U$, that have the smallest cost.  Ties are broken randomly,
which results in slightly different results across different runs.
Once the seed is found, GRAAL builds ``spheres'' of all possible radii
around nodes $v$ and $u$.  A sphere of radius $r$ around node $v$ is
the set of nodes $S_{G}(v,r)=\{x\in V: d(v,x)=r\}$ that are distance
$r$ from $v$ where the distance $d(v,x)$ is the length of the shortest
path from $v$ to $x$.  Spheres of the same radius in two networks are
then greedily aligned together by searching for the pairs $(v',u'):
v'\in S_{G}(v,r)$ and $u'\in S_{H}(u,r)$ that are not already aligned
and that can be aligned with the minimal cost.  When all spheres
around the seed $(v,u)$ have been aligned, some nodes in both networks
may remain unaligned. For this reason, GRAAL repeats the same
algorithm on a pair of networks $(G^{p},H^{p})$ for $p=1,2,$ and $3$,
and searches for the new seed again, if necessary. We define a network
$G^{p}$ as a new network $G^{p}=(V,E^{p})$ with the same set of nodes
as $G$ and with $(v,x)\in E^{p}$ if and only if the distance between
nodes $v$ and $x$ in $G$ is less than or equal to $p$, i.e.,
$d_{G}(v,x)\leq p$.  Note that $G^{1}=G$. Using $G^p, p>1$ allows us
to align a path of length $p$ in one network to a single edge in
another network, which is analogous to allowing ``insertions'' or
``deletions'' in a sequence alignment.  GRAAL stops when each node
from $G$ is aligned to exactly one node in $H$.

GRAAL produces global alignments. We note that optimal global
alignments are not necessarily unique.  Given any particular cost
function, there may be many distinct alignments that all share the
optimal cost.  In this paper, we analyze just one specific alignment
that we believe is a good one, although it may not be optimal even
according to our measure.  Enumerating all optimal (or at least good)
alignments requires extending our algorithm to allow many-to-many
mappings between the nodes in the two networks, and is the subject of
the future work.  Thus, many more predictions of equal validity to
those in this paper are likely to be possible.
However, we empirically demonstrate that a large portion (about 60\%)
of the entire alignment is conserved across different runs of the
algorithm; thus, this core alignment is independent of the randomness
in the algorithm.

The algorithm's pseudo code and details about the complexity analysis
are presented in the Supplementary Information.  The software and data
used in this paper are available upon request.

\subsection{Statistical Significance of our Yeast-Human Alignment}\label{sec:meth3}

Given a GRAAL alignment of two networks $G(V, E)$ and $H(U, F)$, we
compute the probability of obtaining a given or better edge
correctness score at random.
For this purpose, an appropriate null model of random alignment is
required. A random alignment is a random mapping $f$ between nodes in
two networks $G(V, E)$ and $H(U, F)$, $f:V\rightarrow U$. GRAAL
produces \emph{global} alignments, so that all nodes in the smaller
network (smaller in terms of the number of nodes) are aligned with
nodes in the larger network. In other words, $f$ is defined $\forall
v\in V$.  This is equivalent to aligning each edge from $G(V,E)$ with
a \emph{pair of nodes} (not necessarily an edge) in $H(U,F)$. Thus, we
define our null model of random alignment as a random mapping
$g:E\rightarrow U\times U$.
We define $n_{1}=|V|$, $n_{2}=|U|$, $m_{1}=|E|$, and
$m_{2}=|F|$. We also define the number of node pairs in $H$ as $p=\frac{n_{2}(n_{2}-1)}{2}$,
and let $EC=x\%$ be the edge correctness of the given
alignment.  We let $k=[m_{1}\times EC]=[m_{1}\times x]$ be the number of
edges from $G$ that are aligned to edges in $H$.  Then, the
probability $P$ of successfully aligning $k$ or more edges by chance
is the tail of the hypergeometric distribution:
$\textrm{$P$} = \sum_{i=k}^{m_2}\frac{{m_2\choose i} {p - m_2\choose m_1  - i}}{{p \choose m_1}}.$
For our yeast2--human1 alignment, we find $P\approx 7\times 10^{-8}$.

Now we describe how to estimate the statistical significance of the
amount of similarity we find between yeast2 and human1 in our alignment.
To do that, we need to estimate how much similarity one would expect
to find between two {\em random} networks and doing that, in turn,
requires us to specify how we generate model random networks.  Given
two models that purport to fit a set of observations, we generally
consider as superior the one that has fewer tunable parameters.  For
example, the STICKY and ER-DD models are constructed to preserve the
degree distribution of the data.  These and other data-driven models
of random networks
\cite{Throne07,Snijders2002,KuchaievPSB} are thus expected to model
particular PPI networks better than theoretical network models.
However, they are not an appropriate choice to judge whether the
yeast2 and human1 networks share a significant amount of structural
similarity; this is because these models are strongly conditioned on
these particular networks and thus they might transfer onto the model
networks the similarities between yeast2 and human1 that we aim to
detect in the first place. Thus, we search for a well-fitting
\emph{theoretical} null model.
Arguably the best currently known  theoretical model for PPI networks, requiring the
fewest tunable parameters, is the
\emph{geometric random graph} model (``GEO'')
\cite{Przulj04,Przulj2006,Higham2008}, in
which proteins are modeled as existing in a metric space and are
connected by an edge if they are within a fixed, specified distance of
each other.

Although early, incomplete PPI datasets were modeled well by
scale-free networks because of their power-law degree distributions
\cite{Jeong01,Barabasi99}, it has been argued that
such degree distributions were an artifact of
noise\cite{Stumpf05,Vidal05,deSilva2006}. In the light of new PPI
network data, several studies\cite{Przulj04,Przulj2006,Higham2008}
have presented compelling evidence that the structure of PPI networks
is closer to geometric than to scale-free networks.
This was done by comparing frequencies
of graphlets in real-world and model networks \cite{Przulj04}
and by measuring a highly-constraining agreement between ``graphlet degree distributions.''
\cite{Przulj2006} Finally, it has been shown that PPI networks can be successfully
embedded into a low-dimensional Euclidean space, thus directly
confirming that they have a geometric structure \cite{Higham2008}.
The superior fit of the GEO model to PPI networks over other models
may not be surprising, since it can be biologically motivated.  In
particular, the currently accepted paradigm for evolution is based on
a series of gene duplication and mutation events.  We outline our
crude {\it geometric gene duplication} model\cite{GeoGD}. We model
genes, and proteins as their products, as existing in some biochemical
metric space.  Although the dimension and axes of this space are not
obvious, we assume that when a parent gene is duplicated, the child
gene starts at a similar location in the metric space, since it is
structurally identical to the parent and thus inherits interactions
from the parent.  As mutations and ``evolutionary optimization'' act
on the child, it drifts away from the parent in the metric space.  The
child may preserve some of the parent's interacting partners, but it
may also establish new interactions with other genes\cite{GeoGD}.
Similarly, in a geometric graph, the closer two nodes are to each
other, the more interactors they will have in common, and vice-versa.
In addition to PPI networks, GEO is a well-fitting theoretical null
model for other biological networks, e.g., brain function networks
\cite{Kuchaiev2009EMBC} and protein structure networks
\cite{Milenkovic2009}.

Accepting GEO as the optimal null model for PPI networks, we compute
the probability of obtaining the $EC$ of 11.72\% in our alignment of
yeast2 and human1 to be $8.4\times 10^{-3}$. We do so by aligning with
GRAAL pairs of GEO networks of the same size as yeast2 and human1 and
by applying the following form of the Vysochanskij--Petunin
inequality: $P(|X-\mu|\geq\lambda\sigma)\leq\frac{4}{9\lambda^{2}}.$
Since GEO networks that are aligned have \emph{the same} number of
nodes and edges as the data, it is reasonable to assume that the
distribution of their alignment scores is unimodal.  Thus, we use the
Vysochanskij--Petunin inequality, since it is more precise than
Chebyshev's inequality for unimodal distributions.  More details are
supplied in the Supplementary Information.

\section*{Acknowledgments} We thank M. Ra\v{s}ajski for computational assistance.
This project was supported by the NSF CAREER IIS-0644424 grant.

\bibliography{all}

\bibliographystyle{splncs}

\bigskip


\newpage

\newpage

\end{document}